\documentstyle[aps,twocolumn]{revtex}

\renewcommand{\vec}{\bbox}

\newcommand{\be}{\begin{equation}}
\newcommand{\ee}{\end{equation}}
\newcommand{\bea}{\begin{eqnarray}}
\newcommand{\eea}{\end{eqnarray}}
\newcommand{\dd}{\dagger \hspace{-0.7mm} \dagger}
\newcommand{\venus}{ {\scriptscriptstyle +} \hspace{-1.5mm}^{\circ}}
\newcommand{\aeq}{&=&}

\newcommand{\itPi}{{\it \Pi}}
\newcommand{\itSigma}{{\it \Sigma}}

\newcommand{\itOmega}{{\it \Omega}}
\newcommand{\bra}{\langle}
\newcommand{\ket}{\rangle}
\newcommand{\dbra}{\bra \! \bra}
\newcommand{\dket}{\ket \! \ket}

\newcommand{\me}{\mbox{e}}

\begin{document}

\title{Migration of Unstable Vacuum for Dissipative Systems
}

\author{Toshihico Arimitsu\thanks{arimitsu@cm.ph.tsukuba.ac.jp}}

\address{Institute of Physics, University of Tsukuba,
Ibaraki 305-8571, Japan}

\date{\today}

\maketitle

\begin{abstract}

Quantum field theory is constructed upon the assumption of 
stabilities of the vacuum and of the one-particle state. 
For finite temperature, the one-particle state becomes unstable 
because of thermal fluctuations, whereas the thermal vacuum is 
still stable.
In non-equilibrium situation, both the vacuum and 
the one-particle state lose their stability.
Proposed is the introduction of the {\it reference vacuum} which 
takes care of thermal and non-adiabatic time-evolution of a system, 
and produces a time-dependent Fock's representation space.
This may provide us with an extension of the concept of 
{\it dynamical mapping} where a migration among 
unitary inequivalent representation spaces can be handled
for non-equilibrium and dissipative systems.

\end{abstract}


\narrowtext


\section{Introduction}

Quantum field theory is constructed upon a basic assumption
of stability: the stabilities of vacuum and one-particle state.
These stabilities are essential for perturbational calculations
where physical particles are specified by 
asymptotic fields with renormalized masses.
The vacuum, therefore the representation space, is specified 
by the annihilation and creation operators 
of the asymptotic field.
For finite temperature, vacuum keeps its stability but 
one-particle state loses the stability because of 
thermal fluctuations. In this case, we do not have 
an asymptotic field.  We cannot find it out 
by going back to infinite past nor going forward to 
infinite future.
The concept of the {\it dynamical mapping}
\cite{TFD,Umezawa} was introduced to secure this situation, 
where the mapping of Heisenberg operators is not necessarily
performed by means of asymptotic fields.
In non-equilibrium and dissipative systems, the situation 
becomes much worse, where both vacuum and 
one-particle states are unstable.
We do not know how one can extend the concept of 
the dynamical mapping to this challenging situation.

Within Thermo Field Dynamics (TFD) \cite{TFD,Umezawa,Leplae,T-U}, 
two vacuums representing thermal equilibrium states of
different temperatures are 
mutually unitary-inequivalent. 
We can calculate thermodynamic quantities 
by representing corresponding observable operators by means of 
the Fock space constructed on the thermal vacuum
specified by temperature $T$.
Therefore, it is interpreted that the quasi-static process 
induces a change of system among these unitary inequivalent 
representation spaces.

On the other hand, for realistic dynamical processes, 
how can we interpret this migration in the 
set consisting of the orthogonal (inequivalent) representation spaces?
The situation may become more vivid when one remembers 
the process of the vacuum expansion in terms of thermodynamics.
In this paper, by making use of 
Non-Equilibrium Thermo Field Dynamics (NETFD) 
\cite{netfd1,netfd2,netfd3,guida1,guida2,can1,can2,jim,kinetic,%
tft1,hydro,proceedings,stoch,Saito,Zubarev memorial,%
Saito Thesis,Imagire,essay,Sudo,Tominaga-Ban,Ban,Tominaga,%
Iwasaki,Willeboordse,Naoko,non-linear,cloud chamber,Kramers eq},
we will propose a possibility how one can 
deal with the migration of a vacuum among 
the mutually inequivalent representation spaces.
The time-evolution of a thermal vacuum (equivalently the 
time-evolution of a representation space) is controlled by 
dissipative thermal processes, and the dynamics of fields
are specified by mechanical rules.
The former may be described by a macroscopic time scale,
whereas the latter by a microscopic one.

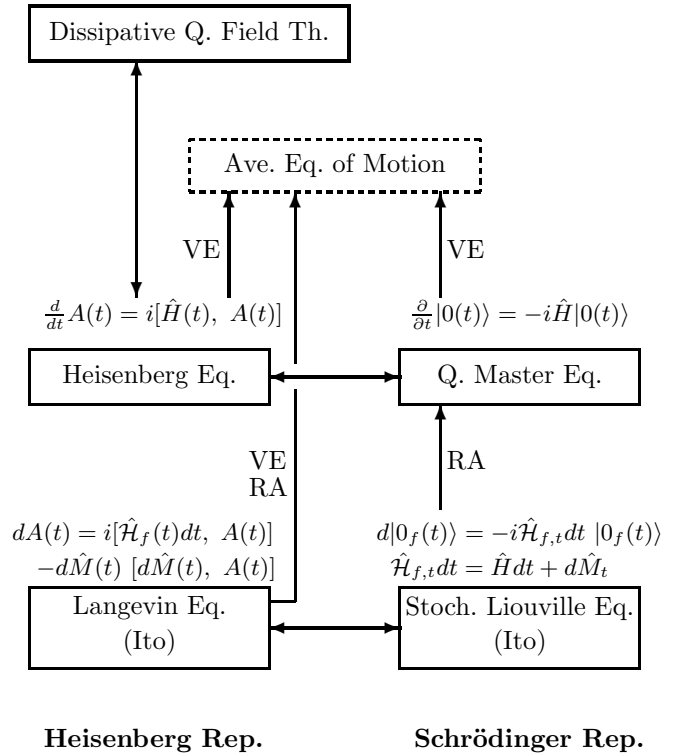
\begin{figure}[h,t,b,p]

\begin{picture}(50,320)(-50,-20)


\thicklines

\put(-45,-10){{\bf Heisenberg Rep.}}
\put(95,-10){{\bf Schr\"odinger Rep.}}

\put(-50,20){\framebox(90,30){}}
\put(-50,22){\makebox(90,15){(Ito)}}
\put(-50,35){\makebox(90,15){Langevin Eq.}}
\put(-40,65){\makebox(65,12.5){\small
$
dA(t) = i [ \hat{{\cal H}}_f(t) dt,\ A(t) ] 
$}}
\put(-30,52){\makebox(55,12.5){\small
$
 - d\hat{M}(t)\  [ d\hat{M}(t),\ A(t) ] 
$}}

\put(90,20){\framebox(90,30){}}
\put(90,22){\makebox(90,15){(Ito)}}
\put(90,35){\makebox(90,15){Stoch.\ Liouville Eq.}}
\put(80,65){\makebox(110,12.5){\small
$
d \vert 0_f(t) \ket = -i \hat{{\cal H}}_{f,t} dt \ \vert 0_f(t) \ket
$}}
\put(80,52){\makebox(95,12.5){\small
$
\hat{{\cal H}}_{f,t} dt = \hat{H} dt + d \hat{M}_t
$}}

\put(90,120){\framebox(90,20){Q.\ Master Eq.}}
\put(80,145){\makebox(110,17.5){\small
$
\frac{\partial}{\partial t} \vert 0(t) \ket = -i \hat{H} \vert 0(t) \ket
$}}

\put(-50,120){\framebox(90,20){Heisenberg Eq.}}
\put(-20,145){\makebox(40,17.5){\small
$
\frac{d}{dt} A(t) = i [\hat{H}(t),\ A(t)]
$}}

\put(10,200){\dashbox{2.0}(110,20){Ave.\ Eq.\ of Motion}}

\put(-50,250){\framebox(120,20){Dissipative Q.\ Field Th.}}

\put(105,80){\vector(0,1){40}}
\put(107.5,95){RA}

\put(50,45){\line(0,1){80}}
\put(50,135){\vector(0,1){65}}
\put(50,45){\line(-1,0){10}}
\put(32.5,95){VE}
\put(32.5,85){RA}

\put(40,35){\vector(1,0){50}}
\put(90,35){\vector(-1,0){50}}

\put(105,160){\vector(0,1){40}}
\put(107.5,175){VE}

\put(25,160){\vector(0,1){40}}
\put(7.5,175){VE}

\put(90,130){\vector(-1,0){50}}
\put(40,130){\vector(1,0){50}}

\put(-10,160){\vector(0,1){90}}
\put(-10,250){\vector(0,-1){90}}

\end{picture}

\caption{System of the Stochastic Differential 
Equations within Non-Equilibrium Thermo Field Dynamics. 
RA stands for the random average. 
VE stands for the vacuum expectation.
}
\label{structure}
\end{figure}

NETFD is a {\it canonical operator formalism} of 
quantum systems in far-from-equili\-brium state which provides us 
with a unified formulation for dissipative systems 
(covering whole the aspects in non-equilibrium statistical 
mechanics, represented by the Boltzmann, the Fokker-Planck, the
Langevin and the stochastic Liouville equations) by the method 
similar to the usual quantum field theory that
accommodates the concept of the dual structure in the
interpretation of nature, i.e.\ in terms of 
the {\it operator algebra} and the {\it representation space}
(Fig.\ \ref{structure}).
The representation space of NETFD (named {\it thermal space}) 
is composed of the direct product of two
Hilbert spaces, the one for {\it non-tilde} fields and the other
for {\it tilde} fields.

The infinitesimal time-evolution generator (hat-Hamiltonian)
of the quantum master equation within NETFD was discovered first 
\cite{netfd1,netfd2} for the cases corresponding to 
stationary processes by, so to speak, a {\em principle 
of correspondence} which makes the connection between NETFD 
and the density operator formalism \cite{Haake,tcl,general tcl}.
Then, it was found  \cite{netfd3} that the time-evolution generator 
can be also derived upon several axioms such as 
(\ref{H hat tildian}) and (\ref{left zero}) below.  
The renormalized time-evolution generator in 
the interaction representation (the semi-free hat-Hamiltonian) 
corresponding to non-stationary processes was derived together with 
an equation for the one-particle distribution function 
\cite{guida1,guida2}.  
Within these aspects,
the canonical formalism of dissipative quantum fields in NETFD was 
formulated, and the close 
structural resemblance between NETFD and usual quantum field 
theories was revealed \cite{can1,can2}.  The generating 
functional within NETFD was derived \cite{jim}. Furthermore, 
the kinetic equation was derived within NETFD \cite{kinetic}, 
and the relation between NETFD 
and the closed time-path methods \cite{schwinger,keldysh,su} was 
shown (see Appendix \ref{Path-Int} for the relation of NETFD to
the path integral method).  
The extension of NETFD to the hydrodynamical region, as well
as the kinetic region, was started
\cite{tft1,hydro,ZT}.

The framework of NETFD was extended further to take 
account of the quantum stochastic processes 
\cite{proceedings,stoch,Saito,Zubarev memorial,Saito Thesis,Imagire}.  
Here again NETFD allowed us to construct a unified canonical theory 
of quantum stochastic operators.  The stochastic Liouville
equations both of the Ito and of the Stratonovich types
were introduced in the Schr\"odinger representation. 
Whereas, the Langevin equations both of the Ito and of the
Stratonovich types were constructed as the Heisenberg
equation of motion with the help of the time-evolution generator 
of corresponding stochastic Liouville equations.  The Ito
formula was generalized for quantum systems.

NETFD has been applied to various systems, e.g.\
the dynamical rearrangement of thermal vacuum in 
superconductor \cite{Sudo}, spin relaxation \cite{Tominaga-Ban},
various transient phenomena in quantum optics 
\cite{Ban,Tominaga,Iwasaki,Willeboordse,Naoko}, 
non-linear damped harmonic oscillator \cite{non-linear},
the tracks in the cloud chamber 
(a non-demolition continuous measurement) \cite{cloud chamber},
microscopic derivation of the quantum Kramers equation 
\cite{Kramers eq}.

\section{Framework of NETFD}

The dynamics of physical systems is described, within
NETFD, by the Schr\"odinger equation for 
the thermal ket-vacuum $\vert 0(t) \ket$:
\be
\frac{\partial}{\partial t} \vert 0(t) \ket 
= -i \hat{H} \vert 0(t) \ket.
\label{Schr eq}
\ee
The time-evolution generator $\hat{H}$ is an tildian operator
satisfying
\be
\left( i \hat{H} \right)^\sim = i \hat{H}.
\label{H hat tildian}
\ee
The {\it tilde conjugation} $\sim$ is defined by 
\cite{Leplae,T-U}: 
\bea
(A_1A_2)^{\sim} \aeq \tilde{A}_1\tilde{A}_2,\\ 
(c_1A_1+c_2A_2)^{\sim} \aeq c^{*}_1\tilde{A}_1+c^{*}_2\tilde{A}_2,
\\
(\tilde{A})^{\sim} \aeq A,\\ 
(A^{\dagger})^{\sim} \aeq \tilde{A}^{\dagger},
\eea
where $c_1$ and $c_2$ are $c$-numbers.
The tilde and non-tilde operators at an equal time are
mutually commutative:
\be
[A,\ \tilde{B}]=0.
\ee

The thermal bra-vacuum $\bra 1 \vert $ is the eigen-vector
of the hat-Hamiltonian $\hat{H}$ with zero eigen-value:
\be
\bra 1 \vert \hat{H} = 0.
\label{left zero}
\ee
This guarantees the conservation of the inner product 
between the bra and ket vacuums in time:
\be
\bra 1 \vert 0(t) \ket = 1.
\label{normalization}
\ee

Let us assume that the thermal vacuums satisfy
\be
\bra 1 \vert^\sim = \bra 1 \vert, \quad 
\vert 0(t_0) \ket^\sim = \vert 0(t_0) \ket,
\ee
at a certain time $t = t_0$. 
Then, (\ref{H hat tildian}) guarantees that 
they are satisfied for all the time:
\be
\bra 1 \vert^\sim = \bra 1 \vert, \quad 
\vert 0(t) \ket^\sim = \vert 0(t) \ket.
\ee

The observable operator $A$ should be an Hermitian operator
consisting only of non-tilde operators. Its expectation 
value is real as can be proven as follows.
\bea
\bra A \ket_t^* \aeq \bra A \ket_t^\sim
\nonumber\\
\aeq \left\{ \bra 1 \vert A \vert 0(t) \ket 
\right\}^\sim 
\nonumber\\
\aeq \left\{ \bra 1 \vert \right\}^\sim \tilde{A} 
\left\{ \vert 0(t) \ket \right\}^\sim
\nonumber\\
\aeq \bra 1 \vert \tilde{A} \vert 0(t) \ket
\nonumber\\
\aeq \bra 1 \vert A^\dagger \vert 0(t) \ket
\nonumber\\
\aeq \bra 1 \vert A \vert 0(t) \ket
\nonumber\\
\aeq \bra A \ket_t.
\eea

\subsection*{An Example --Quantum Kramers Equation}

Let us find out the general structure of hat-Hamiltonian 
which is bilinear in $(x,\ p,\ \tilde{x},\ \tilde{p})$.
$x$ and $p$ satisfies the canonical commutation relation 
\be
[x,\ p] = i.
\ee
Accordingly, $\tilde{x}$ and $\tilde{p}$ satisfies 
\be
[\tilde{x},\ \tilde{p}] = -i.
\ee

The conditions, $(i \hat{H})^\sim = i \hat{H}$, and
$\bra 1 \vert \hat{H} = 0$ give us the general expression
\be
\hat{H} = \hat{H}_0 + i \hat{\itPi},
\ee
where
\be
\hat{H}_0 = H_0 - \tilde{H}_0, \quad 
H_0 = \frac{1}{2m} p^2 + \frac{m \omega^2}{2} x^2,
\ee
\be
\hat{\itPi} = \hat{\itPi}_R + \hat{\itPi}_D,
\ee
with
\bea
\hat{\itPi}_R \aeq - i \frac{1}{2} \kappa \left( x - \tilde{x} \right)
\left( p + \tilde{p} \right),
\nonumber\\
\hat{\itPi}_D \aeq - \frac{1}{2} \kappa m \omega (1+2\bar{n})
\left( x - \tilde{x} \right)^2.
\eea
Here, we neglected the diffusion in $x$-space.
The Schr\"odinger equation 
\be
\frac{\partial}{\partial t} \vert 0(t) \ket = -i \hat{H}
\vert 0(t) \ket,
\label{q Kramers eq}
\ee
gives the {\it quantum Kramers equation}.

The Heisenberg equation for the dissipative system is 
given by
\bea
\frac{d}{dt}x(t) \aeq i [ \hat{H}(t),\ x(t) ]
\nonumber\\
\aeq \frac{1}{m} p(t)+ \frac12 
\kappa \left\{ x(t) - \tilde{x}(t) \right\},
\label{H eq x}
\\
\frac{d}{dt}p(t) \aeq i [ \hat{H}(t),\ p(t) ]
\nonumber\\
\aeq - m\omega^2 x(t) 
- \frac{1}{2} \kappa \left\{ p(t) + \tilde{p}(t) \right\}
\nonumber\\
&&+ i \kappa m \omega 
(1+2\bar{n}) \left\{ x(t) - \tilde{x}(t) \right\}.
\label{H eq p}
\eea

Applying the bra-vacuum $\bra 1 \vert$ of the relevant system,
we have the equations for the vectors:
\bea
\frac{d}{dt}\bra 1 \vert x(t) \aeq \frac{1}{m} \bra 1 \vert p(t),
\nonumber\\
\frac{d}{dt} \bra 1 \vert p(t) \aeq - m\omega^2 \bra 1 \vert x(t) 
- \kappa \bra 1 \vert p(t).
\eea

The stochastic Liouville equation within the Ito calculus becomes
\be
d \vert 0_f(t) \dket = -i \hat{{\cal H}}_{f,t} dt 
\vert 0_f(t) \dket,
\label{q stoch Kramers eq}
\ee
with the stochastic hat-Hamiltonian 
\be
\hat{{\cal H}}_{f,t} dt = \hat{H} dt + d\hat{M}_t.
\ee
Here, the martingale operator $d\hat{M}_t$ is defined by
\be
d\hat{M}_t = \frac{\sqrt{\kappa m \omega}}{2} 
\left( x - \tilde{x} \right) \left(
dX_t + d\tilde{X}_t \right),
\ee
with 
\be
dX_t = dB_t + dB_t^\dagger,
\ee
where $dB$, $dB^\dagger$ and their tilde conjugates are 
the operators representing quantum Brownian motion (see
Appendix~\ref{q Brownian motion}).
The generalized fluctuation-dissipation theorem is given by
\be
d\hat{M}_t d\hat{M}_t = -2\hat{\itPi}_D dt.
\ee
Taking a random average, the stochastic Liouville equation 
(\ref{q stoch Kramers eq}) reduces to 
the quantum master equation (\ref{q Kramers eq})
with 
\be
\vert 0(t) \ket = \bra \vert 0_f(t) \dket.
\ee

The stochastic Heisenberg equation (the Langevin equation)
for this hat-Hamiltonian is given by
\bea
d x(t) \aeq i [ \hat{{\cal H}}_{f}(t) dt,\ x(t) ]
- d\hat{M}(t) \ [d\hat{M}(t),\ x(t) ]
\nonumber\\
\aeq \frac{1}{m} p(t) dt + \frac12 
\kappa \left\{ x(t) - \tilde{x}(t) \right\} dt,
\label{stoch H eq x}
\\
d p(t) \aeq - m\omega^2 x(t) dt
- \frac{1}{2} \kappa \left\{ p(t) + \tilde{p}(t) \right\} dt
\nonumber\\
&& - \frac{\sqrt{\kappa m \omega}}{2} \left\{ dX(t)
+ d \tilde{X}(t) \right\}.
\label{stoch H eq p}
\eea

The averaged equation of motion is given by
\bea
\frac{d}{dt} \dbra x(t) \dket \aeq \frac{1}{m} \dbra p(t) \dket,\\
\frac{d}{dt} \dbra p(t) \dket \aeq - m\omega^2 \dbra x(t) \dket
- \kappa \dbra p(t) \dket,
\eea
where $\dbra \cdots \dket = \bra 1 \vert \bra \vert \cdots \vert \ket 
\vert 1 \ket$. The vacuumes $\bra \vert$ and $\vert \ket$ are introduced in 
Appendix \ref{q Brownian motion}.
These averaged equations can be also derived from (\ref{H eq x})
and (\ref{H eq p}) by taking the average $\dbra \cdots \dket$.

\section{Stationary State}

Let us assume the existence of the eigen-equation
\be
\hat{H} \vert 0_E \ket = \hat{E} \vert 0_E \ket,
\label{eigen-equation}
\ee
where the eigen-value $\hat{E}$ is a complex c-number
called hat-energy.

Applying $\bra 1 \vert$ to (\ref{eigen-equation}) and 
making use of (\ref{left zero}) and (\ref{normalization}),
we see that 
\be
0 = \bra 1 \vert \hat{H} \vert 0_E \ket 
= \hat{E} \ \bra 1 \vert 0_E \ket 
= \hat{E}.
\ee
Therefore, if there exists the eigen-vector satisfying
(\ref{eigen-equation}), its eigen-value $\hat{E}$ is 
equal to zero.
The Schr\"odinger equation (\ref{Schr eq}) tells us that
the eigen-vector $\vert 0_E \ket$ is a stationary state.

If we assume that there is only one equilibrium state, 
the stationary eigen-vector represents 
the thermal equilibrium state which will be referred to as
$\vert 0_{eq} \ket$.

\section{Vacuum for Canonical Ensemble (TFD)} \label{ensemble}

Let us consider two sets of boson operators 
$(a,\ a^\dagger)$ and $(\tilde{a},\ \tilde{a}^\dagger)$
which satisfy the commutation relation:
\be
[ a^\mu ,\ \bar{a}^\nu ] = \delta^{\mu,\nu},
\ee
where we introduced the thermal doublet notation defined by
\bea
a^{\mu=1} = a, && \quad a^{\mu=2} = \tilde{a}^{\dagger},
\\
\bar{a}^{\mu=1} = a^{\dagger}, && \quad \bar{a}^{\mu=2} = -\tilde{a},
\eea
and the Kronecker delta $\delta^{\mu \nu}$.
These operators satisfy the thermal state conditions 
\bea
a \vert 0_{eq}(\theta) \ket \aeq \me^{-\omega/T(\theta)} \tilde{a}^\dagger
\vert 0_{eq}(\theta) \ket, 
\label{thermal state cond 1}\\
\bra 1(\theta) \vert a^\dagger \aeq \bra 1(\theta) \vert \tilde{a}.
\label{thermal state cond 2}
\eea
The thermal vacuums $\bra 1(\theta)\vert$ and 
$ \vert 0_{eq}(\theta) \ket$ are
tilde-invariant:
\be
\bra 1(\theta) \vert^\sim = \bra 1(\theta) \vert, \quad 
\vert 0_{eq}(\theta) \ket^\sim = \vert 0_{eq}(\theta) \ket.
\label{tilde inv}
\ee

For simplicity, we will confine the discussion to the case of
Boson fields, which can be easily extended to Fermion fields.

The vacuums $\bra 1(\theta)\vert $ and 
$\vert 0_{eq}(\theta) \ket $ representing the canonical ensemble 
are defined by 
\be
\bra 1(\theta) \vert \tilde{\gamma}_\theta^{\venus} = 0,\quad
\gamma_\theta \vert 0_{eq}(\theta) \ket = 0,
\ee
with the creation operator $\tilde{\gamma}_\theta^{\venus}$ and 
the annihilation operator $\gamma_\theta$ which are 
defined through the Bogoliubov transformation 
\cite{TFD}-\cite{T-U}
\be
\gamma_\theta^\mu = B(\theta)^{\mu\nu} a^\nu,\quad
\bar{\gamma}_\theta^\mu = \bar{a}^\nu B^{-1}(\theta)^{\nu\mu}.
\ee
Here, we introduced the thermal doublet notations
\bea
\gamma_\theta^{\mu=1} = \gamma_\theta, && \quad
\gamma_\theta^{\mu=2} = \tilde{\gamma}^{\venus},
\\
\bar{\gamma}_\theta^{\mu=1} = \gamma^{\venus}, && \quad
\bar{\gamma}_\theta^{\mu=2} = - \tilde{\gamma}_\theta,
\eea
and the matrix
\be
 B(\theta)^{\mu\nu} = \left(
 \begin{array}{cc}
  1+\bar{n}(\theta) & -\bar{n}(\theta) \\
  -1 & 1 \\
 \end{array}
 \right).
\label{Bogoliubov theta matrix}
\ee
Representing, for example, the number operators 
$a^\dagger a $ with respect to this vacuum, we obtain
\be
\bra 1(\theta) \vert a^\dagger a \vert 0_{eq}(\theta) \ket
= \bar{n}(\theta).
\ee
Using the thermal state conditions (\ref{thermal state cond 1}) 
and (\ref{thermal state cond 2}), we see that
\be
\bar{n}(\theta)= \left( e^{\omega/T(\theta)} - 1 \right)^{-1},
\ee
which is the thermal vacuums representing 
the canonical ensemble with temperature $T(\theta)$.
Note that two representation spaces belonging to different 
parameter $\theta$ are unitary-inequivalent 
\cite{TFD,Umezawa,Leplae,T-U}.

Imagine a space constituted by the set of mutually inequivalent 
degenerate thermal vacuums $\vert 0_{eq}(\theta) \ket$ specified by 
$\bar{n}(T(\theta))$ having zero hat-energy, i.e., $\hat{E} = 0$.
The adiabatic change of the thermal equilibrium state 
specified by temperature $T(\theta)$ to the one by $T(\theta')$
can be interpreted as the change of the parameter 
$\theta$ to $\theta'$ in this space.

For non-equilibrium cases, thermal vacuum, which is not the 
zero hat-energy vacuum in general, is specified 
by the one-particle distribution function $n_k(t)$. 
Its different dependence in $k$ gives
mutually inequivalent vacuums.
A participation of certain dissipative process makes it
possible to connect these inequivalent vacuums in time
as will be shown in the following.

\section{Semi-Free Operators}

Now, we consider, for simplicity, the case of a semi-free field
\bea
\varphi(x)^\mu \aeq \int \frac{d^3 k}{(2 \pi)^{3/2}}\ 
e^{i \vec{k} \cdot \vec{x}} a_k(t)^\mu, \\
\bar{\varphi}(x)^\mu \aeq \int \frac{d^3 k}{(2 \pi)^{3/2}}\ 
e^{-i \vec{k} \cdot \vec{x}} \bar{a}_k(t)^\mu,
\eea
accompanied by the equal-time commutation relation
\be
[ a_k(t)^\mu,\ \bar{a}_{\ell}(t)^\nu ] = \delta^{\mu \nu} 
\delta(\vec{k} - \vec{\ell}).
\ee
Here, we introduced the thermal doublet notation 
for the semi-free operators by
\bea
a_k(t)^{\mu=1} = a_k(t), && \quad
a_k(t)^{\mu=2} = \tilde{a}_k^{\dd}(t),
\\
\bar{a}_k(t)^{\mu=1} = a_k^{\dd}(t), && \quad
\bar{a}_k(t)^{\mu=2} = -\tilde{a}_k(t).
\eea

The Heisenberg operators $a_k(t)^\mu$ and $\bar{a}_k(t)^\nu$ 
are defined by
\be
a_k(t)^\mu = \hat{V}^{-1}(t) a_k^\mu \hat{V}(t),\quad
\bar{a}_k(t)^\mu = \hat{V}^{-1}(t) \bar{a}_k^\mu \hat{V}(t),
\ee
where $\hat{V}(t)$ satisfies 
\be
 \frac{d}{dt} \hat{V}(t) = -i \hat{H}_t \hat{V}(t),
\quad \left( i \hat{H}_t \right)^\sim = i \hat{H}_t,
\label{eq for V-hat}
\ee
with $\hat{V}(0)=1$, i.e.,
\be
 a_k(t) = \hat{V}^{-1}(t) a_k \hat{V}(t),\quad 
 \tilde{a}_k^{\dd}(t)= \hat{V}^{-1}(t) \tilde{a}_k^\dagger \hat{V}(t).
\ee
Note that $a_k$ and $\tilde{a}_k^\dagger$ do not depend on time.
Since the semi-free hat-Hamiltonian $\hat{H}_t$
is not necessarily Hermite, we introduced the symbol $\dd$
in order to distinguish it from the Hermite conjugation 
$\dagger$.
However in the following, we will use $\dagger$ instead of $\dd$, 
for simplicity, unless it is confusing.
We also drop the subscript
representing momentum unless it is necessary.

It is known that the hat-Hamiltonian for the renormalized 
semi-free field is given by \cite{guida1,guida2}
\be
\hat{H}_t =  \hat{H}_{0,t} + i \hat{\itPi}_{t},
\label{H hat semi-free}
\ee
with
\bea
\hat{H}_{0,t} \aeq \omega(t) \bar{a}^\mu a^\mu
+ \omega(t), 
\label{H0-hat}\\
\hat{\itPi}_{t} \aeq - \bar{a}^\mu \left[
\kappa(t) A(t)^{\mu\nu} + \frac{d n(t)}{dt} \tau^{\mu\nu} 
\right] a^\nu  + \kappa(t),
\label{H1-hat}
\eea
where the integration with respect to the momentum $k$ is 
implicit.
We introduced two matrices
\be
 A(t)^{\mu\nu} = \left(
 \begin{array}{cc}
  1+2n(t) & -2n(t) \\
  2(1+n(t)) & -(1+2n(t)) \\
 \end{array}
 \right),
\label{A matrix}
\ee
and, \(\tau^{11} = \tau^{21} = 1,\ \tau^{12} = \tau^{22} = -1\).
Here, $n(t)$ represents the one-particle distribution function 
which should be specified by the kinetic equation
\be
\frac{d}{dt} n(t) = -2 \kappa(t) n(t) +i \itSigma^<(t),
\label{kinetic eq}
\ee
where the functions $\kappa(t)$ and $\itSigma^<(t)$ 
($\omega(t)$ also) are determined 
self-consistently when an interaction Hamiltonian is specified.

The equation of motion for the field is given by
the Heisenberg equation
\be
\frac{d}{dt} \varphi(x)^\mu = i [ \hat{H}(t),\ \varphi(x)^\mu ],
\ee
with
\be
\hat{H}(t) = \hat{V}^{-1}(t) \hat{H}_t \hat{V}(t).
\ee
Then, the equation of motion for the semi-free operator $a(t)^\mu$ 
becomes
\bea
\frac{d}{dt} a(t)^\mu \aeq -i \left[ \omega(t) \delta^{\mu \nu} 
- i \kappa(t) A(t)^{\mu \nu} \right] a(t)^\nu
\nonumber\\
&& - \frac{dn(t)}{dt} \tau^{\mu \nu} a(t)^\nu.
\eea

Introducing annihilation and creation operators
\bea
\gamma_t^{\mu=1} = \gamma_t, && \quad
\gamma_t^{\mu=2} = \tilde{\gamma}^{\venus},
\\
\bar{\gamma}_t^{\mu=1} = \gamma^{\venus}, && \quad
\bar{\gamma}_t^{\mu=2} = 
- \tilde{\gamma}_t,
\eea
in the Schr\"odinger representation through
\begin{equation}
 \gamma_t^\mu = B(t)^{\mu\nu} a^\nu,\quad 
\bar{\gamma}_t^\mu =  \bar{a}^\nu B^{-1}(t)^{\nu\mu},
\label{Bogoliubov transformation}
\end{equation}
with the {\em time-dependent Bogoliubov transformation}
\be
 B(t)^{\mu\nu} = \left(
 \begin{array}{cc}
  1+n(t) & -n(t) \\
  -1 & 1 \\
 \end{array}
 \right),
\label{Bogoliubov matrix}
\ee
the hat-Hamiltonians (\ref{H0-hat}) and
(\ref{H1-hat}) reduce, respectively, to \cite{guida1,guida2}
\bea
\hat{H}_{0,t} \aeq \int d^3 k \omega_k(t) 
\left( \gamma_k^{\venus} \gamma_{k,t} - \tilde{\gamma}_k^{\venus}
 \tilde{\gamma}_{k,t} \right), 
\label{H0-hat normal}\\
\hat{\itPi}_{t} \aeq - \int d^3 k\ \left[
\kappa_k(t) \left( \gamma_k^{\venus} \gamma_{k,t} 
+ \tilde{\gamma}_k^{\venus}
 \tilde{\gamma}_{k,t} \right) \right. 
\nonumber\\
&&
\left. 
- \frac{dn_k(t)}{dt} \gamma_k^{\venus}
 \tilde{\gamma}_k^{\venus} \right].
\label{H1-hat normal}
\eea
The new operators annihilate the thermal vacuums
$\vert 0(t) \ket$ and $\bra 1 \vert$ as
\be
 \gamma_{k,t} \vert 0(t) \ket \ = 0,\quad \bra1\vert 
\tilde{\gamma}_k^{\venus} = 0,
\ee
respectively.
The time-dependence of $\gamma_t$ and
$\tilde{\gamma}_t$ comes from that of the one-particle 
distribution function $n(t)$:
\be
\gamma_t = \gamma_{t=0} - \left[n(t) - n(0) \right] 
\tilde{\gamma}^{\venus}.
\label{gamma sub t}
\ee

Substituting the normal ordered expression of the hat-Hamiltonian
(\ref{H0-hat normal}) and (\ref{H1-hat normal}) 
into the Schr\"odinger equation 
(\ref{Schr eq}), we can obtain its solution as
\bea
\vert 0(t) \ket \aeq \hat{V}(t) \vert 0 \ket 
\nonumber\\
\aeq \exp \left\{ \int d^3 k
\left[n_k(t)-n_k(0) \right] \gamma_k^{\venus} 
 \tilde{\gamma}_k^{\venus} \right\} \vert0\ket.
\label{condensation of pairs}
\eea
This expression tells us that the time-evolution of 
the unstable vacuum is realized by the condensation of 
$\gamma_k^{\venus}  \tilde{\gamma}_k^{\venus}$-pairs
into the vacuum.
It also shows that the vacuum is the functional of
the one-particle distribution function $n_k(t)$.
The dependence of the thermal vacuum on $n_k(t)$ is
given by
\be
\frac{\delta}{\delta n_k(t)} \vert 0(t) \ket = 
\gamma_k^{\venus} \tilde{\gamma}_k^{\venus} \vert 0(t) \ket.
\ee
We see that the vacuum $\vert 0(t) \ket$ represents the state
where exists the macroscopic object described by 
the one-particle distribution function $n_k(t)$.

The operators
\be
\gamma(t)^\mu = \hat{V}^{-1}(t) \gamma_t^\mu \hat{V}(t),
\quad
\bar{\gamma}(t)^\mu =  \hat{V}^{-1}(t) \bar{\gamma}_t^\mu \hat{V}(t),
\label{gamma t}
\ee
in the Heisenberg representation with the thermal doublets, 
\bea
\gamma(t)^{\mu = 1} = \gamma(t), && 
\gamma(t)^{\mu = 2} = \tilde{\gamma}^{\venus}(t),
\\
\bar{\gamma}(t)^{
\mu = 1} = \gamma^{\venus}(t), && \bar{\gamma}(t)^{\mu = 2} = -
\tilde{\gamma}(t),
\eea
satisfy the equal-time commutation relation:
\be
[\gamma(t)^\mu,\ \bar{\gamma}(t)^\nu ] = \delta^{\mu\nu},
\ee
and have the properties
\begin{equation}
 \gamma(t) \vert 0\ket \ = 0,\quad \bra1\vert 
\tilde{\gamma}^{\venus}(t) = 0.
\label{annihilation of vacuum}
\end{equation}
Note that the thermal vacuums are tilde invariant (see 
(\ref{tilde inv})).

The quasi-particle operators $\gamma(t)^\mu$ 
satisfy the equation of motion
\be
\frac{d}{dt} \gamma(t)^\mu 
=  -i \left[ \omega(t) \delta^{\mu \nu} - i \kappa(t) 
\tau_3^{\mu \nu} \right] \gamma(t)^\nu,
\label{eq for quasi-particle}
\ee
where the matrix \(\tau_3^{\mu \nu}\) is defined by
\(\tau_3^{11} = - \tau_3^{22} = 1,\ \tau_3^{12} = \tau_3^{21} = 0\).
The equation of motion (\ref{eq for quasi-particle}) is solved to give
\be
\gamma(t)^\mu = \exp \left\{ \int_{0}^t\!\!ds \left[ -i\omega
 (s) \delta^{\mu \nu} -\kappa(s) \tau_3^{\mu \nu} \right] 
\right\} \gamma(0)^\nu.
\label{solution of gamma}
\ee


\section{Reference Vacuum}

In order to put the part relating to the non-equilibrium and 
dissipative time-evolution into vacuum, we divide $\hat{V}(t)$
into two parts:
\be
\hat{V}(t) = \hat{V}_0(t) \hat{W}(t),
\ee
where $\hat{V}_0(t)$ satisfies
\be
\frac{d}{dt} \hat{V}_0(t) = -i \hat{H}_{0,t} \hat{V}_0(t).
\ee
Then, we see that $\hat{W}(t)$ obeys
\be
\frac{d}{dt} \hat{W}(t) = \hat{\itPi}(t) \hat{W}(t),
\ee
with
\be
\hat{\itPi}(t) = \hat{V}^{-1}_0(t) \hat{\itPi}_{t} \hat{V}_0(t) 
= \hat{\itPi}_{t},
\ee
where we used the commutativity
\be
[\hat{H}_{0,t},\ \hat{\itPi}_{t}] = 0.
\ee

Operators in this representation evolves in time as
\bea
a(t)^\mu \aeq \hat{V}_0^{-1}(t) a^\mu \hat{V}_0(t)
= a^\mu e^{-i\int_0^t ds\ \omega(s)},\\
\gamma(t)^\mu \aeq \hat{V}_0^{-1}(t) \gamma_t^\mu \hat{V}_0(t)
= \gamma_t^\mu e^{-i\int_0^t ds\ \omega(s)}.
\eea
Here, we are using the same notation $a(t)^\mu$ and 
$\gamma(t)^\mu$ for the quasi-particle operators as those
for the semi-free particle operators introduced in the previous
section. We hope that they are not confusing.

We call the vacuum
\bea
\vert 0(t) \ket_{\mbox{{\footnotesize ref}}} \aeq \hat{W}(t) 
\vert 0 \ket 
\nonumber\\
\aeq \exp \left\{ \int d^3 k
\left[n_k(t)-n_k(0) \right] \gamma_k^{\venus} 
 \tilde{\gamma}_k^{\venus} \right\} \vert 0 \ket,
\label{reference vacuum}\\
 _{\mbox{{\footnotesize ref}}}\bra 1 \vert 
\aeq \bra 1 \vert \hat{V}_0(t) = \bra 1 \vert,
\eea
the {\it reference vacuum} \cite{essay}.
For the latter equation, we used the fact that
\be
\bra 1 \vert \hat{H}_{0,t} = 0.
\ee
The Schr\"odinger equation for the former vacuum 
$\vert 0(t) \ket_{\mbox{{\footnotesize ref}}}$
can be rewritten as
\be
\left\{ \frac{\partial}{\partial t} 
+ \int d^3k\ \frac{d n_k(t)}{d t} \frac{\delta}{\delta n_k(t)} \right\}
\vert 0(t) \ket_{\mbox{{\footnotesize ref}}}
= 0.
\label{migration of vacuum}
\ee
This shows that the reference vacuum, in this case, is migrating 
in the super-representation space spanned by 
the one-particle distribution function $\{ n_k(t) \}$
with the {\it velocity} $\{ d n_k(t)/dt \}$ as a conserved quantity.
These vacuums satisfy
\be
_{\mbox{{\footnotesize ref}}}\bra 1 \vert \tilde{\gamma}^{\venus}(t)
=0,\quad
\gamma(t) \vert 0(t) \ket_{\mbox{{\footnotesize ref}}} = 0.
\ee

For the case of a semi-free particle corresponding to 
a {\it stationary process} \cite{netfd1,netfd2}
where
\be
i \itSigma^<(t) = 2\kappa \bar{n},\quad \omega(t)=\omega,
\quad \kappa(t)= \kappa,
\ee
the kinetic equation becomes
\be
\frac{d n(t)}{dt} = 2 \kappa \left[ n(t) - \bar{n} \right].
\ee
Then, the hat-Hamiltonian (\ref{H hat semi-free}) reduces to 
\be
\hat{H} = \hat{H}_0 + i \hat{\itPi},
\label{b2}
\ee
with
\be
\hat{H}_0 = H_0 - \tilde{H}_0,\quad
H_0 = \omega a^\dagger a,
\ee
\bea
\hat{\itPi} \aeq - \kappa \left[ \left(1+2\bar{n}\right) \left( a^\dagger 
 a + \tilde{a}^\dagger \tilde{a} \right) \right.
\nonumber\\
&& \left. - 2\left(1+\bar{n}\right)
 a \tilde{a} - 2\bar{n} a^\dagger \tilde{a}^\dagger \right] 
 - 2\kappa\bar{n}.
\eea
The time-dependence of the reference vacuum is given by
\be
\vert 0(t) \ket_{\mbox{{\footnotesize ref}}} = \exp \left\{ 
\left[ \bar{n} - n(0) \right] \left( 1 - e^{-2 \kappa t}
\right)  \gamma^{\venus} 
\tilde{\gamma}^{\venus} \right\}  \vert 0 \ket,
\label{simple ref vacuum}
\ee
where $\bar{n}$ is the one-particle distribution function of the 
final equilibrium state, say $n(\theta)$ in section \ref{ensemble}.
For this simple case, $\vert 0(t) \ket$ is determined by 
$\bar{n}$, $n(0)$ and $\kappa$.
It is interesting if we put $n(0) = n(\theta_0)$
and $\bar{n} = n(\theta)$, and remember 
that, in non-equilibrium thermodynamics, the thermodynamical states
are specified only by its initial and final states.

The representation space (the thermal space) of NETFD is
the vector space spanned by the set of bra and ket state
vectors which are generated, respectively, by cyclic
operations of the annihilation operators $\gamma(t)$ and 
$\tilde{\gamma}(t)$ on 
$_{\mbox{{\footnotesize ref}}}\bra 1 \vert$, and of the creation
operators $\gamma^{\venus}(t)$ and
$\tilde{\gamma}^{\venus}(t)$ on 
$\vert 0(t) \ket_{\mbox{{\footnotesize ref}}}$.

The normal product is defined by means of the annihilation
and the creation operators, i.e.\ $\gamma^{\venus}(t),\ 
\tilde{\gamma}^{\venus}(t)$ stand to the left of
$\gamma(t),\ \tilde{\gamma}(t)$. 
The process, rewriting physical operators in terms of the
annihilation and creation operators, leads to a Wick-type
formula, which in turn leads to Feynman-type diagrams for
multi-point functions in the renormalized interaction
representation.  The internal line in the Feynman-type
diagrams is the unperturbed two-point function (the propagator):
\bea
 G(t,t')^{\mu\nu} \aeq -i \bra1\vert T \left[a(t)^\mu \bar{a}(t')^\nu 
  \hat{W}(\bullet) \right] \vert 0  \ket_{\mbox{{\footnotesize ref}}}  
\nonumber\\
\aeq \left[ B^{-1}(t) {\cal G}(t,t') B(t') \right]^{\mu\nu},
\label{two point func}
\eea
where
\begin{eqnarray}
 {\cal G}(t,t')^{\mu\nu} \aeq -i \bra1\vert T 
\left[ \gamma(t)^\mu \bar{\gamma}
 (t')^\nu  \hat{W}(\bullet) \right] \vert 0 \ket_{\mbox{{\footnotesize ref}}}  
\nonumber\\
\aeq \left(
  \begin{array}{cc}
   G^R(t,t') & 0 \\
   0 & G^A(t,t') \\
  \end{array}
 \right),
\label{a6}
\end{eqnarray}
with
\begin{eqnarray}
\lefteqn{ G^R(t,t') }\nonumber\\
\aeq -i \theta(t-t') \exp \left\{ \int_{t'}^t\!\!ds \left[ -i\omega
 (s)-\kappa(s) \right] \right\}, \\
\lefteqn{ G^A(t,t') }\nonumber\\
\aeq i \theta(t'-t) \exp \left\{ \int_{t'}^t\!\!ds \left[ -i\omega
 (s)+\kappa(s) \right] \right\}.
\end{eqnarray}
Here, the time argument $\bullet$ represents a time which is larger than
$t$ and $t'$.

\section{Dynamical Mapping}

Within the reference vacuum representation, we have
\bea
a(t) \aeq \gamma(t) + n(t) \tilde{\gamma}^{\venus}(t),
\\
\tilde{a}^\dagger(t) \aeq \left\{1+n(t) \right\}
\tilde{\gamma}^{\venus}(t) + \gamma(t),
\eea
which can be interpreted as the dynamical mapping of 
the operators of $a(t)$ and $\tilde{a}^\dagger(t)$ with respect to
the reference vacuum $_{\mbox{{\footnotesize ref}}}\bra 1 \vert$ 
and $\vert 0(t) \ket_{\mbox{{\footnotesize ref}}}$.
The fact that the reference vacuum is specified by 
the one-particle distribution function $n(t)$ is shown in 
the dynamical mapping 
\bea
a^\dagger(t) a(t) \aeq n(t) + \left\{1+n(t) \right\} 
\gamma^{\venus}(t) \gamma(t)
\nonumber\\
&& + n(t) \tilde{\gamma}^{\venus}(t) \tilde{\gamma}(t) 
\nonumber\\
&& + n(t) \left\{1+n(t) \right\} \gamma^{\venus}(t) 
\tilde{\gamma}^{\venus}(t)
+ \gamma(t) \tilde{\gamma}(t).
\eea

The dynamical mapping of the fluctuation is given by
\bea
\lefteqn{a^\dagger(t) a(t) a^\dagger(t) a(t) - 
_{\mbox{{\footnotesize ref}}}\bra 1 \vert a^\dagger(t) a(t) \vert 0(t) 
\ket_{\mbox{{\footnotesize ref}}}^2
}\nonumber\\
\aeq n(t) \left\{1+n(t) \right\} + [\mbox{normal ordered series}].
\eea

\section{An Interpretation of the Reference Vacuum Representation}
\label{RV}

In order to understand the physical meaning of 
the reference vacuum, we will consider, for a moment, 
the Schr\"odinger equation 
\be
\frac{\partial}{\partial t} \vert 0(t) \ket = 
-i \left( \hat{H}_0 + \epsilon \hat{H}_1 \right) \vert 0(t) \ket,
\ee
with $\epsilon \ll 1$.

Let us suppose that the vacuum $\vert 0(t) \ket$ depends on
two {\it independent} time variables $t_0$ and $t_1$:
\be
\vert 0(t) \ket \rightarrow \vert 0(t_0, t_1) \ket,
\ee
where $t_0$ represents a fast (or microscopic) time scale
and $t_1$ a slow (or macroscopic) time scale.
The {\it slow} and {\it fast} in time scale 
is introduced with respect to 
the small parameter $\epsilon$.
Then, the time derivative will be given by
\be
\frac{\partial}{\partial t} = \frac{\partial}{\partial t_0} 
+ \epsilon \frac{\partial}{\partial t_1}.
\ee
Physical axis can be provided by putting $t_0=t$ and 
$t_1 = \epsilon t$ at an appropriate stage.

Expanding the vacuum with respect to $\epsilon$:
\be
\vert 0(t_0, t_1) \ket = \vert 0(t_0, t_1) \ket_0
+ \epsilon \vert 0(t_0, t_1) \ket_1,
\ee
we obtain the equations of order of $\epsilon^0$ and 
of $\epsilon^1$, respectively, in the forms
\be
\frac{\partial}{\partial t_0} \vert 0(t_0, t_1) \ket_0 = 
-i \hat{H}_0 \vert 0(t_0, t_1) \ket_0,
\label{ep 0}
\ee
\bea
\lefteqn{\frac{\partial}{\partial t_0} \vert 0(t_0, t_1) \ket_1 
+ i \hat{H}_0 \vert 0(t_0, t_1) \ket_1 
}\nonumber\\
\aeq - \frac{\partial}{\partial t_1} \vert 0(t_0, t_1) \ket_0 
- i \hat{H}_1 \vert 0(t_0, t_1) \ket_0.
\label{ep 1}
\eea
The solution of (\ref{ep 0}) is given by
\be
\vert 0(t_0, t_1) \ket_0 = \me^{-i \hat{H}_0 t_0} 
\vert 0(t_1) \ket_0.
\label{sol 0}
\ee
Note that the vacuum $\vert 0(t_1) \ket_0$ appeared on 
the right-hand side depends only on the slow time $t_1$.

Substituting (\ref{sol 0}) into the right-hand side of 
(\ref{ep 1}), we have inhomogeneous terms which should be
vanished unless they give rise to secular divergence.
Then, we obtain the equation which determines the $t_1$-dependence
of $\vert 0(t_1) \ket_0$ in the form
\be
\frac{\partial}{\partial t_1} \vert 0(t_1) \ket_0 
= -i \hat{H}_1(t_0) \vert (t_1) \ket_0,
\label{ep 0a}
\ee
with
\be
\hat{H}_1(t_0) = \me^{i \hat{H}_0 t_0} \hat{H}_1 
\me^{-i \hat{H}_0 t_0}.
\ee
The solution 
\be
\vert 0(t_1) \ket_0 = \me^{-i\hat{H}_1(t_0) t_1} \vert 0 \ket_0,
\ee
of (\ref{ep 0a}) gives the time-dependence of the reference vacuum.
This slow time-dependence may be related to that of macroscopic (or 
semi-macroscopic) objects. The point is that we put this 
slow time-dependent part into one of the characteristics of 
the reference vacuum.

Since the inhomogeneous terms in (\ref{ep 1}) has been taken out, 
the equation for $\vert 0(t_0, t_1) \ket_1$ reduces to
\be
\frac{\partial}{\partial t_0} \vert 0(t_0, t_1) \ket_1 
= -i \hat{H}_0 \vert 0(t_0, t_1) \ket_1.
\ee
The vacuum $\vert 0(t_1) \ket_1$ which appears in the solution
of this equation is determined by the secular terms in 
the equation of order of $\epsilon^2$.

It is necessary to determine the dynamics with respect to 
the fast time variable $t_0$ and the one with respect to 
the slow time variable $t_1$
self-consistently in the sense that the thermalization of 
the reference vacuum and the microscopic time-evolution become
mutually consistent on the physical axis.
There are similar interpretations supporting this point of view.
The time-evolution of a vacuum specified by 
the stochastic Liouville equation 
is controlled by a random process whose time index is 
semi-macroscopic \cite{stoch,Saito,Zubarev memorial}.

Note that the separation of time scale into two is intimately 
related to the Boltzmann equation limit.
If this limit is not appropriate, as in the hydrodynamical regime,
one needs to introduce more independent time variables
which may leads to multiple time-scale analysis \cite{Hansen1,Hansen2}.
In the hydrodynamical regime, 
the vacuum describing non-equilibrium thermodynamic 
processes is a functional of the space and time dependent
thermodynamic quantities, e.g., the local temperature,
the local pressure, the local chemical potential, 
the local fluid velocity and so on \cite{hydro}.
The space and time indices of these quantities are 
macroscopic.

\section{Comments and Future Problems}

We introduced in this paper one of the new concepts revealed by 
the development of NETFD, the concept of migration of unstable vacuum,
which is inevitable for the dissipative quantum systems in far-from 
equilibrium states. The concept was clearly stated by the expression 
(\ref{migration of vacuum}) which shows that in the kinetic regime
the reference vacuum 
migrates with the velocity $dn_k(t)/dt$ as a conserved quantity
in the super-representation space spanned by 
the one-particle distribution function $n_k(t)$.
In order to understand the migration among inequivalent representation 
spaces, it is required to extend the concept of dynamical mapping 
for dissipative non-equilibrium systems, which was initiated in this 
paper.
In this respect, an extension of the mechanism and of the concept in 
the appearance of macroscopic objects
due to the Boson transformation to non-equilibrium dissipative 
situations is one of the attractive future problems.
It is also necessary to extend the present approach to cope with 
phenomena in the hydrodynamical regime \cite{hydro}, 
such as dense fluid, dense nuclear matter, dense plasma and so on.

The formulation of NETFD has a wide potentiality 
in the sense that it may provide us with a technical tool 
telling how to extend 
the concepts revealed in the usual quantum field theory to dissipative 
non-equilibrium cases. 
The discovery that the time-evolution of 
unstable vacuum is realized by the condensation of 
$\gamma_k^{\venus}  \tilde{\gamma}_k^{\venus}$-pairs
into the vacuum is one of the examples.
We hope that with the help of NETFD we can find out a breakthrough to realize 
Boltzmann's original dream, i.e., the essential understanding of 
irreversibility.
The notion of the dual structure in quantum field theory, 
the operator algebra and the representation space, was not recognized 
in Boltzmann's days.

\acknowledgements

The authors would like to thank Dr.~N.~Arimitsu, 
Dr.~T.~Saito, Dr.~A.~Tanaka, Mr.~T.~Imagire, Mr.~T.~Indei 
and Mr.~Y.~Endo for their collaboration with fruitful discussions.

\appendix

\section{Relation to Path Integral}\label{Path-Int}

The kernel 
\be
K(\alpha, \beta,t; \alpha', \beta',t') = (\alpha, \beta \vert
\me^{-i \hat{H} (t-t')} \vert \alpha', \beta' ).
\ee
defined by
\bea
\lefteqn{
(\alpha, \beta \vert 0(t) \ket 
}\nonumber\\
\aeq \int \frac{d^2 \alpha'}{\pi}
\int \frac{d^2 \beta'}{\pi} K(\alpha, \beta,t; \alpha', \beta',0)
(\alpha',\beta' \vert 0 \ket
\eea
can be expressed in terms of path integral.
Here, we introduced the coherent state by
\be
\vert \alpha, \beta ) =
\mbox{e}^{\alpha a^{\dagger} -\alpha^{*} a} 
\mbox{e}^{\beta^{*} \tilde{a}^{\dagger} -\beta \tilde{a}} 
\vert 0, \tilde{0} ), 
\ee
with
\be
a \vert \alpha,\beta ) = \alpha 
\vert \alpha, \beta ),\quad
\tilde a \vert \alpha,\beta ) = \beta^{*} \vert 
\alpha, \beta )
\ee
\be
( \alpha,\beta \vert a^\dagger = 
( \alpha,\beta \vert \alpha^{*},\quad
( \alpha,\beta \vert \tilde{a}^{\dagger} = 
( \alpha,\beta \vert \beta.
\ee

By making use of 
\bea
\lefteqn{
(\alpha_{n},\beta_{n} 
\vert \alpha_{n-1},\beta_{n-1} )
}\nonumber\\
\aeq \mbox{e}^{-\frac{1}{2} \vert \alpha_{n} \vert^{2}
 -\frac{1}{2}\vert \alpha_{n-1} \vert^{2} + \alpha^{*}_{n} 
 \alpha_{n-1}}
\nonumber\\
&& \times \mbox{e}^{-\frac{1}{2} \vert \beta_{n-1} \vert^{2} 
-\frac{1}{2}\vert \beta_{n} \vert^{2} + \beta^{*}_{n-1} \beta_{n}}
\eea
and
\bea
\lefteqn{h_{n,n-1} ( \alpha_{n},\beta_{n} \vert \alpha_{n-1},\beta_{n-1} )
}\nonumber\\
\aeq  ( \alpha_{n},\beta_{n} 
\vert \hat H \vert
\alpha_{n-1},\beta_{n-1} )  \nonumber \\
\aeq \left\{ \omega \alpha_{n}^{*} \alpha_{n-1}-\omega 
\beta_{n} \beta_{n-1}^{*}    \right. \nonumber\\
&& \left. - i\kappa \left[(1+2\bar{n})
(\alpha_{n}^{*} \alpha_{n-1} + \beta_{n} \beta_{n-1}^{*}) 
\right. \right. 
 \nonumber \\
   & & -\left. \left. 2(1+\bar{n})\alpha_{n-1}\beta_{n-1}^{*}
    - 2\bar{n}\alpha_{n}^{*}\beta_{n} \right] -2i\kappa \bar{n} 
\right\} \nonumber \\
   && \times 
( \alpha_{n},\beta_{n} \vert \alpha_{n-1},\beta_{n-1} )
\eea
the path integral 
\bea
\lefteqn{K(\alpha,\beta,t;\alpha',\beta',t') } 
\nonumber\\
\aeq \lim_{N \to \infty} \left( \prod\limits_{i = 1}^{N-1} 
   \int \frac{d^{2}\alpha_{i}}{\pi} \right)
                     \left( \prod\limits_{j = 1}^{N-1} 
   \int \frac{d^{2}\beta_{j}}{\pi} \right)
\nonumber\\
&& \quad \quad \times
( \alpha_{N},\beta_{N} \vert \alpha_{N-1},\beta_{N-1} ) 
\cdots 
( \alpha_1,\beta_1 \vert \alpha_0,\beta_0 ) 
\nonumber\\
&& \quad \quad \times \mbox{exp} \left[-i \Delta t 
\sum_{n=1}^{N-1} h_{n,n-1} \right] 
\nonumber \\
\aeq \ \ \mbox{e}^{-\frac{1}{2} \vert \alpha \vert^{2}-\frac{1}{2}
\vert \alpha' \vert^{2}-\frac{1}{2} \vert \beta \vert^{2}
+ \frac{1}{2} \vert \beta' \vert^{2} } 
\int {\cal D}^2 \alpha \int {\cal D}^2 \beta 
\nonumber \\
& \times & \!\!\! \mbox{exp} 
\left[ i\! \int_{t'}^{t} dt
\left[ i \alpha^{*}(t) \dot{\alpha}(t) 
- i  \beta^{*}(t)\dot{\beta}(t)
\right. \right.
\nonumber \\
&& -\alpha(t) \left\{ \omega \alpha^{*}(t)
-i\kappa(1+2\bar{n}) \alpha^{*}(t)
+2 i \kappa(1+\bar{n})\beta^{*}(t) \right\}
\nonumber \\
&& -\beta(t) \left\{-\omega \beta^{*}(t)
- i \kappa(1+2\bar{n})\beta^{*}(t)
+2 i \kappa \bar{n} \alpha^{*}(t) \right\}
\nonumber \\
&& \left. \left.
-2\kappa \bar{n}  \right] \right].
\eea
can be solved with the help of the equations of motion
\be
\frac{d}{dt} \left(
\begin{array}{c}
          \alpha^{*}(t) \\
          \beta^{*}(t)
\end{array}
\right)
= \itOmega
         \left(
         \begin{array}{c}
         \alpha^{*}(t) \\
         \beta^{*}(t) 
         \end{array}
         \right).
\ee
\be
\itOmega =     \left(
         \begin{array}{cc}
         \omega-i\kappa(1+2\bar{n}) & 2i\kappa (1+\bar{n}) \\
         -2i\kappa \bar{n}          & \omega+i\kappa(1+2\bar{n})
         \end{array} 
         \right)
\ee
with the boundary conditions
\bea
\alpha^{*}(t)=\alpha^{*},\quad  \alpha(t')=\alpha' 
\nonumber\\ 
\beta^{*}(t')= \beta^{*},\quad \beta(t)=\beta.
\eea

The result is 
\bea
\lefteqn{K(\alpha, \beta,t; \alpha', \beta',t') }\nonumber\\
\aeq \nu(t-t') \ \mbox{exp} 
\left[-\frac{1}{2}{\vert \alpha \vert}^{2}-\frac{1}{2}{\vert\alpha' \vert}^{2}-\frac{1}{2}{\vert \beta \vert}^{2}-\frac{1}{2}
{\vert \beta' \vert}^{2} \right. \nonumber\\ 
&& + k_i(t-t')\ \beta'^{*} \alpha' 
+ k_f(t-t')\ \alpha^{*} \beta
\nonumber\\ 
&& + \left. \ell(t-t')\ \alpha^{*} \alpha'
+ \ell^*(t-t')\ \beta'^{*} \beta
\right]
\eea
where
\bea
k_i(t) \aeq (1+\bar{n})\nu(t) \ (1-\mbox{e}^{-2\kappa t})
\\
k_f(t) \aeq \bar{n} \nu(t) \ (1-\mbox{e}^{-2\kappa t})
\\
\ell(t) \aeq \nu(t) \ \me^{-i\omega t- \kappa t}
\\
\nu(t) \aeq \frac{1}{(1+\bar{n})-\bar{n} \mbox{e}^{-2\kappa t}}.
\eea

Then, we finally obtain
\bea
\lefteqn{
(\alpha, \beta \vert 0(t) \ket 
}\nonumber\\
\aeq \int \frac{d^2 \alpha'}{\pi}
\int \frac{d^2 \beta'}{\pi} K(\alpha, \beta,t; \alpha', \beta',0)
(\alpha',\beta' \vert 0 \ket
\nonumber\\
\aeq \frac{1}{1+n(t)}\ \me^{-\frac{1}{2}{\vert \alpha \vert}^{2}
-\frac{1}{2}{\vert \beta \vert}^{2} + \frac{n(t)}{1+n(t)}
\alpha^* \beta }
\eea

\section{Quantum Brownian Motion} \label{q Brownian motion}

Let us introduce the annihilation and creation operators 
$b_t$, $b^\dagger_t$ and their tilde conjugates 
satisfying the canonical commutation relation:
\be
[ b_t,\ b^\dagger_{t'} ] = \delta(t - t'),\quad 
[ \tilde{b}_t,\ \tilde{b}^\dagger_{t'} ] = \delta(t - t').
\ee
The vacuums $( \vert$ and $\vert )$ are defined by 
\be
b_t \vert ) = 0,\quad \tilde{b}_t \vert ) = 0,\quad 
( \vert b^\dagger_t = ( \vert \tilde{b}_t.
\ee
The argument $t$ represents time.

Introducing the operators
\bea
B_t \aeq \int_0^{t-dt} dB_{t'} = \int_0^t dt'\ b_{t'},\\
B^\dagger_t \aeq \int_0^{t-dt} dB^\dagger_{t'} 
= \int_0^t dt'\ b^\dagger_{t'},
\label{B}
\eea
and their tilde conjugates for $t \geq 0$, we see that they satisfy
$
B(0) = 0
$, 
$
B^\dagger(0)=0
$, 
\be
[B_s,\ B^\dagger_t ] = \mbox{min}(s,t),
\label{commutation B}
\ee
and their tilde conjugates, and that 
they annihilate the vacuum $\vert )$ with the thermal state condition for 
$( \vert$:
\be
dB_t \vert ) = 0,\quad d\tilde{B}_t \vert ) = 0,\quad
( \vert d B^\dagger_t = ( \vert d\tilde{B}_t.
\ee
These operators represent the quantum Brownian motion.

Let us introduce a set of new operators by the relation
\be
dC_t^\mu = {\cal B}^{\mu \nu} dB_t^\nu,
\ee
with the Bogoliubov transformation defined by
\bea
 {\cal B}^{\mu\nu} = \left(
  \begin{array}{cc}
   1+ \bar{n} & -\bar{n} \\
   -1 & 1 \\
  \end{array}
 \right),
\label{B bar}
\eea
where $\bar{n}$ is the Planck distribution function.
We introduced the thermal doublet:
\bea
dB_t^{\mu=1} = dB_t, && \quad dB_t^{\mu=2} = d\tilde{B}^\dagger_t,
\\
d\bar{B}_t^{\mu =1} = dB^\dagger_t, && \quad d\bar{B}_t^{\mu = 2} = -
d\tilde{B}_t,
\eea
and the similar doublet notations for $dC_t^\mu$ and 
$d{\bar C}_t^\mu$.
The new operators annihilate the new vacuum $\bra \vert$, and 
have the thermal state condition for $\vert \ket$:
\be
dC_t \vert  \ket = 0,\quad d\tilde{C}_t \vert  \ket = 0, 
\quad 
\bra  \vert d C^\dagger_t = \bra \vert 
d\tilde{C}_t.
\label{cal B bra}
\ee

We will use the representation space constructed on
the vacuums $\bra \vert$ and $\vert \ket$.  
Then, we have, for example,
\bea
\bra \vert dB_t \vert \ket \aeq \bra \vert dB^\dagger_t \vert \ket = 0,\\
\bra \vert dB^\dagger_t dB_t \vert \ket = \bar{n} dt,
&&
\bra \vert dB_t dB^\dagger_t \vert \ket 
= \left( \bar{n}+1 \right) dt.
\label{B-Bd}
\eea



\begin{references}

\bibitem{TFD} H.~Umezawa, H.~Matsumoto and M.~Tachiki, {\it
        Thermo Field Dynamics and Condensed States}
        (North-Holland 1982).
\bibitem{Umezawa} H.~Umezawa, {\it Advanced Field Theory 
	--- Micro, Macro, and Thermal Physics ---}
	(American Institute of Physics 1993), and the references 
	therein.
\bibitem{Leplae} L.\ Leplae, H.\ Umezawa and F.\ Mancini,
	Physics Reports {\bf 10C}, 153 (1974).
\bibitem{T-U} Y.\ Takahashi and H.\ Umezawa, 
	Collect.\ Phenom.\ {\bf 2}, 55 (1975).
\bibitem{netfd1} T.~Arimitsu and H.~Umezawa, 
	Prog.\ Theor.\ Phys.\ {\bf 74}, 429 (1985).
\bibitem{netfd2} T.~Arimitsu and H.~Umezawa, 
	Prog. Theor. Phys. {\bf 77}, 32 (1987).
\bibitem{netfd3} T.~Arimitsu and H.~Umezawa, 
	Prog. Theor. Phys. {\bf 77}, 53 (1987).
\bibitem{guida1} T.~Arimitsu, M.~Guida and H.~Umezawa, 
	Europhys. Lett. {\bf 3}, 277 (1987).
\bibitem{guida2} T.~Arimitsu, M.~Guida and H.~Umezawa, 
	Physica {\bf A148}, 1 (1988).
\bibitem{can1} T.~Arimitsu and H.~Umezawa, 
	J. Phys. Soc. Japan {\bf 55}, 1475 (1986).
\bibitem{can2} T.~Arimitsu, H.~Umezawa and Y.~Yamanaka, 
	J. Math. Phys. {\bf 28}, 2741 (1987).
\bibitem{jim} T.~Arimitsu, J.~Pradko and H.~Umezawa, 
	Physica {\bf A135}, 487 (1986).
\bibitem{kinetic} T.~Arimitsu, 
	Physica {\bf A148}, 427 (1988).
\bibitem{tft1} T.~Arimitsu, 
	Physica {\bf A158} (1989) 317.
\bibitem{hydro} T.~Arimitsu, 
	J.~Phys.~A: Math.~Gen.\ {\bf 24}, L1415 (1991).
\bibitem{proceedings} T.~Arimitsu, 
	{\it Thermal Field Theories}, eds.\ H.~Ezawa,
        T.~Arimitsu and Y.~Hashimoto (North-Holland, 1991) 207.
\bibitem{stoch} T.~Arimitsu,
	Phys.\ Lett.\ {\bf A153}, 163 (1991).
\bibitem{Saito}T.~Saito and T.~Arimitsu, 
	Modern Phys.\ Lett.\ B {\bf 6}, 1319 (1992).
\bibitem{Zubarev memorial} T.~Arimitsu, Condensed Matter Physics 
	(Lviv, Ukraine) {\bf 4} (1994) 26, and the references therein.
\bibitem{Saito Thesis}	T.~Saito and T.~Arimitsu,
	J.\ Phys.\ A: Math.\ Gen.\ {\bf 30}, 7573 (1997).
		T.~Saito and T.~Arimitsu.
\bibitem{Imagire}  T.~Imagire, T.~Saito, K.~Nemoto and T.~Arimitsu,
		Physica A {\bf 256}, 129 (1997).
\bibitem{essay} T.~Arimitsu,
		Physics Essays {\bf 9}, 591 (1996).
\bibitem{Sudo} T. Arimitsu, Y. Sudo and H. Umezawa,
           Physica {\bf 146A}, 433 (1987).
\bibitem{Tominaga-Ban} T. Tominaga, M. Ban, T. Arimitsu, 
		J. Pradko and H. Umezawa.
            Physica {\bf 149A}, 26 (1988).
\bibitem{Ban} M. Ban and T. Arimitsu,
            Physica {\bf 146A}, 89 (1987).
\bibitem{Tominaga} T. Tominaga, T. Arimitsu, J. Pradko and H. Umezawa,
           Physica {\bf A150}, 97 (1988).
\bibitem{Iwasaki} T.~Iwasaki, T.~Arimitsu and F.H.~Willeboordse,
            {\it Thermal Field Theories}, ed.~H.~Ezawa, T.~Arimitsu and 
            Y.~Hashimoto (North-Holland 1991) 459.
\bibitem{Willeboordse} T.~Arimitsu, F.H.~Willeboordse and T.~Iwasaki,
           Physica A {\bf 182}, 214 (1992).
\bibitem{Naoko} T.~Arimitsu and N.~Arimitsu,
               Phys.\ Rev.\ E {\bf 50}, 121 (1994).
\bibitem{non-linear} T.~Saito and T.~Arimitsu,
               Mod.\ Phys.\ Lett.\ B {\bf 7}, 623 (1993).
\bibitem{cloud chamber}	T.~Arimitsu,
		Proceedings of the {\it International Conference on Stochastic 
		Processes and Their Applications} at Chennai (Madras) in India,
		(1998) in press.
\bibitem{Kramers eq} T.~Satio and T.~Arimitsu,
   		Proceedings of the {\it International Conference on Stochastic 
		Processes and Their Applications} at Chennai (Madras) in India,
		(1998) in press.
\bibitem{Haake} F.~Haake, 
	{\it Springer Tracts in Modern
        Physics}, vol.\ 66 (Springer-Verlag, 1973) 98.
\bibitem{tcl} F.~Shibata and T.~Arimitsu, 
	J. Phys. Soc. Japan {\bf 49}, 891 (1980).
\bibitem{general tcl} T.~Arimitsu, 
	J. Phys. Soc. Japan {\bf 51}, 1720 (1982).
\bibitem{schwinger} J.~Schwinger, 
	J. Math. Phys. {\bf 2}, 407 (1961).
\bibitem{keldysh} L.~V.~Keldysh, 
	Sov. Phys. JETP {\bf 20}, 1018 (1965).
\bibitem{su} K.~Chou, Z.~Su, B.~Hao and L.~Yu, 
	Phys. Rep. {\bf 118}, 1 (1985).
\bibitem{ZT} Zubarev and Tokarchuk admired the method of NETFD, 
	and they also started to
	use it for the investigation of these regions \cite{Zubarev} 
	(see also \cite{Tokarchuk} for the application to the problem of 
	the quark-gluon plasma).
\bibitem{Zubarev} D.~N.~Zubarev and M.~V.~Tokarchuk,
        Theor.\ Math.\ Phys.\ {\bf 88}, 876 (1991).
\bibitem{Tokarchuk} M.~V.~Tokarchuk, 
	(1993) Lviv, preprint (ICMP-93-7E).
\bibitem{Hansen1} L.~Bocquet, J.~Piasecki and J-P.~Hansen,
	J.\ Stat.\ Phys.\ {\bf 76}, 505 (1994).
\bibitem{Hansen2} L.~Bocquet, J-P.~Hansen and J.~Piasecki,
	J.\ Stat.\ Phys.\ {\bf 76}, 527 (1994).


\end{references}
\end{document}